\documentclass[10pt,twocolumn,superscriptaddress,floats,showpacs,nobalancelastpage,longbibliography,prb]{revtex4-2}

\usepackage{array,longtable}
\newcolumntype{L}{>{\tiny $}p{0.33\columnwidth}<{$}}
\newcolumntype{M}{>{\scriptsize $}p{0.33\columnwidth}<{$}}
\newcolumntype{N}{>{\scriptsize $}p{0.43\columnwidth}<{$}}
\usepackage{amsmath}
\usepackage{amssymb}
\usepackage{amsfonts}
\usepackage{graphicx}
\usepackage{hyperref}
\usepackage{tabularx}
\usepackage{upgreek}

\usepackage{blindtext}

\usepackage{subfiles} 

\usepackage{float}
\usepackage{graphics}
\usepackage[caption=false]{subfig}
\usepackage{tikz}
\usepackage{times}
\usepackage{color}

\usepackage[normalem]{ulem}

\allowdisplaybreaks[1]

\newif\ifhyper
\hypertrue
\ifhyper
\hypersetup{
   citecolor = {green},
   colorlinks = {true}, 
   urlcolor = {blue}, 
   linkcolor = {blue}
}
\fi

\begin{document}

\title{Field-induced Bose-Einstein condensation and supersolid in the two-dimensional Kondo necklace}

\author{Wei-Lin Tu}
\affiliation{Division of Display and Semiconductor Physics, Korea University, Sejong 30019, Korea}

\author{Eun-Gook Moon}
\affiliation{Department of Physics, Korea Advanced Institute of Science and Technology (KAIST), Daejeon 34141, Korea}

\author{Kwan-Woo Lee}
\affiliation{Division of Display and Semiconductor Physics, Korea University, Sejong 30019, Korea}
\affiliation{Department of Applied Physics, Graduate School, Korea University, Sejong 30019, Korea}

\author{Warren E. Pickett}
\email{wepickett@ucdavis.edu}
\affiliation{Department of Physics, University of California, Davis, California 95616, USA}

\author{Hyun-Yong Lee}
\email{hyunyong@korea.ac.kr}
\affiliation{Division of Display and Semiconductor Physics, Korea University, Sejong 30019, Korea}
\affiliation{Department of Applied Physics, Graduate School, Korea University, Sejong 30019, Korea}
\affiliation{Interdisciplinary Program in E$\cdot$ICT-Culture-Sports Convergence, Korea University, Sejong 30019, Korea}

\date{\today}

\begin{abstract}
The application of an external magnetic field of sufficient strength to a spin system composed of a localized singlet can overcome the energy gap and trigger bosonic condensation and so provide an alternative method to realize exotic phases of matter in real materials. Previous research has indicated that a spin Hamiltonian with on-site Kondo coupling may be the effective many-body Hamiltonian for $\text{Ba}_2\text{NiO}_2\text{(AgSe)}_2$ (BNOAS) and here we study such a Hamiltonian using a tensor network ansatz in two dimensions. Our results unveil a phase diagram which indicates the underlying phases of BNOAS. We propose, in response to the possible doping-induced superconductivity of BNOAS, a fermionic model for further investigation. We hope that our discovery can bring up further interest in both theoretical and experimental researches for related nickelate compounds.
\end{abstract}

\maketitle

\section{\label{Intro}Introduction}

Bose-Einstein condensation\,(BEC) is a phenomenon where a finite fraction of the quasiparticles in the system condenses into a single quantum mechanical entity on a macroscopic scale as a consequence of quantum statistical effects~\cite{Bose24, Einstein24}. It leads to an exotic phase of matter, superfluid, which is a fluid with zero viscosity. The superfluid was originally discovered in the liquid helium-4~\cite{Kapitza38}. On the other hand, it was found that the fraction of BEC is strongly suppressed due to the strong interaction between $^4$He atoms. Since then, great effort has been devoted to the search for weakly interacting or dilute Bose gases. 

One of the most promising platforms is the quantum magnet~\cite{Batista04} which hosts many bosonic excitations such as magnon, triplon, and spinon. Especially, one can tune the boson density by applying a magnetic field to induce condensation. Indeed, magnetic field-induced order is a widely studied phenomenon~\cite{nphys893, RevModPhys.86.563}. Experimentally, spin-singlet compounds such as $\text{TlCuCl}_3$~\cite{nature423, LORENZ2007291, doi:10.1143/JPSJ.77.013701, PhysRevB.77.134420, ncomms12822} and $\text{SrCu}_2\text{(BO}_3\text{)}_2$ (well-known for its field-induced solid orders)~\cite{PhysRevB.62.15067,PhysRevLett.100.090401, PhysRevLett.111.137204, Haravifard14372, arxiv:2107.02929} with $\text{S}=1/2$ spin singlets, and $\text{CsFeBr}_3$ with $\text{S}=1$ singlets~\cite{PhysRevB.96.144404} have demonstrated such effects under magnetic field. 

Theoretical studies~\cite{PhysRevLett.84.5868, PhysRevLett.89.077203, PhysRevB.69.054423, PhysRevLett.102.060602, PhysRevB.80.224419} also unveiled the underlying mechanism for such effects. In a spin-singlet material, the local singlet state serves as the ground state with a finite gap, separating itself from triplet excitons. Upon applying a magnetic field, the three-fold excited states split into three triplon bands with $S^z$=$+1, 0, -1$. The branch with spins aligning along the external field becomes soft and gradually reduces the gap to complete closure~\cite{nature423, doi:10.1143/JPSJ.66.1900}. After the gap closing, a condensation of triplons that breaks $\text{U}(1)$ symmetry takes place, leading to the effective spin superfluid (SSF) phase.

For such effects to take place, clearly we need a spin-singlet phase as the precursor. A recent study by Jin {\it et al.}~\cite{PhysRevResearch.2.033197} of a magnetic material with layered nickelate $\text{Ba}_2\text{NiO}_2\text{(AgSe)}_2$ (BNOAS), recently synthesized under high pressure and high temperature~\cite{bnoas_exp}, proposed the origin of its peculiar susceptibility $\chi_{\text{sp}}$.  
It is constant above 150 K and the same constant below 110 K in zero field, with a peak at T$^*$=130 K. Thus there are no free moments at high or low temperature, yet a magnetic reconstruction occurs at 130 K giving the peak in $\chi_{\text{sp}}$. 

This behavior can be explained as arising from local spin singlets (contributing nothing to susceptibility) at high and low T, with some reconstruction occurring at T$^*$~\cite{PhysRevResearch.2.033197}.
Correlated first principles calculations predict a ground state consisting of a novel singlet within the Ni $\text{e}_\text{g}$ subshell, made of spins with the local ``Kondo-like" spin texture: the $\text{d}_{\text{x}^2-\text{y}^2}$ electron (or hole, depending on viewpoint) is coupled with the $\text{d}_{\text{z}^2}$ electron (or hole) to an unusual spin singlet with internal orbital structure and highly anisotropic exchange coupling~\cite{PhysRevResearch.2.033197}, with the first signature of such an on-site, orbital entangled Lee-Pickett singlet having been seen in calculations on the infinite layer nickelate LaNiO$_2$~\cite{LP2004prb}. Due to the above-mentioned facts, in the studies of BNOAS' quantum effect one might need to focus on the area where spin singlet state serves as the ground state.

Jin {\it et al.}~\cite{PhysRevResearch.2.033197} proposed an effective spin Hamiltonian, named the Kondo sieve, for describing the spin behavior of BNOAS:
\begin{equation}
\begin{aligned}
H_{\text{KS}}=&J\sum_{\langle i,j \rangle}\boldsymbol{\sigma}_i \cdot \boldsymbol{\sigma}_j+K\sum_{i}\boldsymbol{\sigma}_i \cdot \boldsymbol{\tau}_i\\
&+J_z\sum_{\lbrack i,j \rbrack}\boldsymbol{\tau}_i \cdot \boldsymbol{\tau}_j-\sum_{i}\boldsymbol{S}_i \cdot \boldsymbol{h}\\
=&H_{\text{KN}}+J_z\sum_{\lbrack i,j \rbrack}\boldsymbol{\tau}_i \cdot \boldsymbol{\tau}_j-\sum_{i}\boldsymbol{S}_i \cdot \boldsymbol{h}.
\end{aligned}
\label{Kondo-sieve}
\end{equation}
The local spin moment $\boldsymbol{\sigma}_i$ (from the $\text{d}_{\text{x}^2-\text{y}^2}$ orbital of BNOAS) is Kondo-coupled with the $\boldsymbol{\tau}_i$ spin moment (from the same-site $\text{d}_{\text{z}^2}$ orbital) with exchange coupling $K$. $\boldsymbol{\sigma}_i$ and $\boldsymbol{\tau}_i$ fulfill the commutation relation that $[\xi^a,\xi^b]=\text{i}\epsilon_{abc}\xi^c$ with $\xi=\sigma$ or $\xi=\tau$, and $\epsilon_{abc}$ being the Levi-Civita symbol. Within a layer, for the nearest neighbor, denoted by $\langle i,j \rangle$, $\boldsymbol{\sigma}$ moments are coupled by the Heisenberg $J$ term. Due to the multilayered nature of BNOAS, neighboring NiO$_2$ layers have $J_z$ coupling between $\boldsymbol{\tau}$-spin neighbors $[i,j]$ along $\hat{\text{z}}$. The $i$-site total spin operator $\boldsymbol{S}_i=\boldsymbol{\sigma}_i+\boldsymbol{\tau}_i$ is coupled with the external field through a Zeeman field $\boldsymbol{h}$. Note that the model enjoys global $\text{U}(1)$ symmetry with $\text{e}^{\text{i}\alpha \sum_i S_i^z}$, for total spin and with an arbitrary $\alpha$.
Eq. (\ref{Kondo-sieve}) as a whole stands for the Kondo sieve model $H_{\text{KS}}$, while its first part, for each layer, represents the Kondo-necklace model $H_{\text{KN}}$ in two dimensions (2D)~\cite{DONIACH1977231}. 
For the on-site $K$ term, we have $\langle\psi_\text{singlet}|K\boldsymbol{\sigma}_i \cdot \boldsymbol{\tau}_i|\psi_\text{singlet}\rangle=-\frac{3}{4}K$ and $\langle\psi_\text{triplet}|K\boldsymbol{\sigma}_i \cdot \boldsymbol{\tau}_i|\psi_\text{triplet}\rangle=\frac{1}{4}K$.
Therefore, an on-site singlet-triplet splitting with magnitude equal to $K$ takes place.
This system provides a good platform, by introducing an external magnetic field $\boldsymbol{h}$, for field-induced BEC phases.

After gap closing due to the magnetic field, the mechanism can be thought of as a bosonic system, which makes it intriguing to consider the possibility of hosting a state with coexisting diagonal and off-diagonal orders, the so-called spin supersolid state~\cite{PhysRev.104.576}. For the usual hard-core Bose-Hubbard-like Hamiltonian on a square lattice, one needs more than the nearest-neighbor interaction to stabilize the spin supersolid phase~\cite{PhysRevLett.84.1599, PhysRevB.65.014513}. Various spin supersolid phases can be detected by introducing frustration~\cite{PhysRevB.96.045119, JPhysCondensMatter.32.455401} or dipole-dipole interaction~\cite{PhysRevLett.104.125301, doi:10.1143/JPSJ.80.113001, PhysRevLett.112.127203, PhysRevA.86.063635, NewJPhys.17.123014, PhysRevA.102.053306, PhysRevA.103.043333}. Related researches for $\text{{S}r{C}u}_2\text{(B}\text{O}_3\text{)}_2$ which can be addressed by the Shastry-Sutherland model~\cite{PhysRevB.62.15067, PhysRevLett.100.090401, PhysRevLett.111.137204, arxiv:2107.02929} or spin dimers~\cite{PhysRevB.78.184418, PhysRevB.88.014419} also indicated the formation of spin supersolid.
However, despite the recently proposed spin supersolid phase induced by the spin-orbital coupling~\cite{nature.543.91, PhysRevLett.120.140403, PhysRevA.100.031602}, due to the difficulty of experimental realization such spin supersolid states cannot be easily observed. On the other hand, field-induced condensation might provide a better platform for such exotic phases. It has been shown previously that with such a two-spin Hamiltonian, by breaking the $\text{SU}(2)$ symmetry with an anisotropic strength $\Delta$ that controls the coupling of $\sigma^z_i\sigma^z_j$, formation of a spin supersolid can be triggered by external field~\cite{PhysRevB.73.014433, NG20071371, PhysRevLett.99.027202}. In this work, we study the Kondo necklace model $H_{\text{KN}}$ using a 2D tensor network ansatz called infinite projected entangled-pair states (iPEPS)~\cite{PhysRevLett.101.250602}. As one will see, our results reveal that not only the BEC but also spin supersolid can be triggered by the magnetic field, suggesting a good potential of BNOAS for studying exotic phases of matter. 

\begin{figure*}[pbt]
\centering
	\includegraphics[width=2.0\columnwidth]{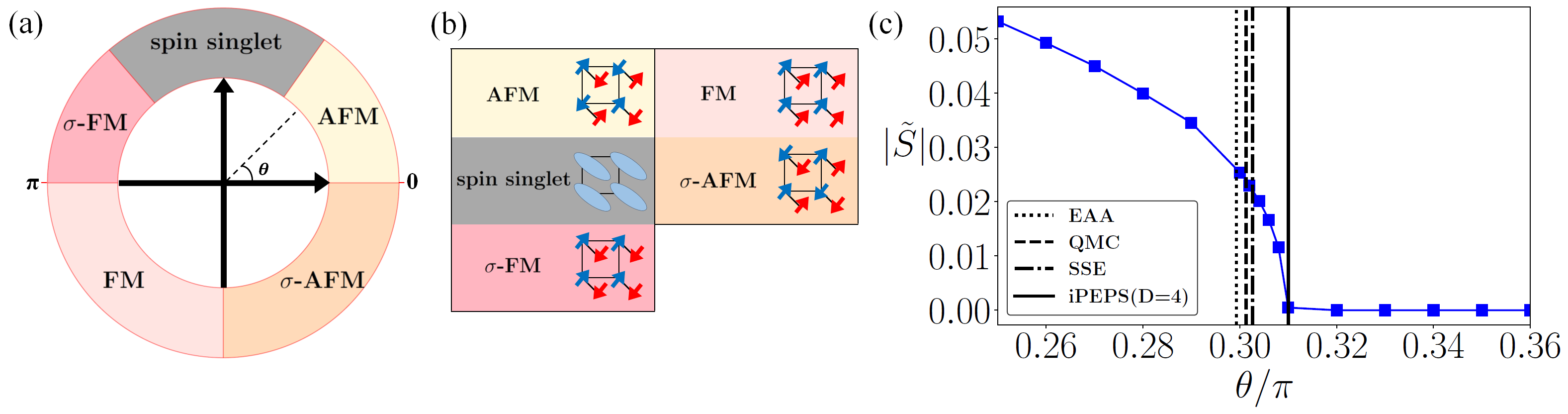}
\caption{Infinite projected entangled-pair state (iPEPS) results at zero field. (a) Phase diagram versus $\theta=$tan$^{-1}(K/J)$ (for details, see Eqs.~(\ref{Hamiltonian}) and (\ref{Hamiltonianxxz})) of 2D Kondo-necklace model ($\Delta=1.0$, given in Eq.~(\ref{Hamiltonianxxz})) ground states without external field, demonstrated as a hollowed pie chart. Starting from $\theta=0$, we have in sequence the antiferromagnetic (AFM), spin singlet, $\sigma$-ferromagnetic ($\sigma$-FM), ferromagnetic (FM), and $\sigma$-antiferromagnetic ($\sigma$-FM) phases as the ground state. The schematic spin configuration of each phase is illustrated in (b). Blue arrows indicate $\boldsymbol{\sigma}$, while red arrows are the Kondo spin moments $\boldsymbol{\tau}$. Light blue ovals represent the local spin singlet on each site. (c) Plot of $|\tilde{S}|$ order parameter (Eq.~(\ref{Sorder})) along with $\theta$. We also mark the phase transition points obtained by three other methods, effective analytical approach (EAA), quantum Monte Carlo (QMC), and stochastic series expansion (SSE), for comparison.}
\label{fig1}
\end{figure*} 

\section{\label{result}Results}
\subsection{\label{zerofield}Ground States in Zero Field}

For discussing the phase diagram, we first generalize to the 2D Kondo-necklace XXZ model in the following way:
\begin{equation}
\begin{aligned}
H= \text{cos}\theta\sum_{\langle i,j \rangle}(\boldsymbol{\sigma}_i \cdot \boldsymbol{\sigma}_j)_{\Delta}+\text{sin}\theta\sum_{i}\boldsymbol{\sigma}_i \cdot \boldsymbol{\tau}_i-h\sum_{i}S^z_i,
\end{aligned}
\label{Hamiltonian}
\end{equation}
where
\begin{equation}
\begin{aligned}
&J=\text{cos}\theta, K=\text{sin}\theta,\\
&(\boldsymbol{\sigma}_i \cdot \boldsymbol{\sigma}_j)_{\Delta}=\sigma^x_i\sigma^x_j+\sigma^y_i\sigma^y_j+\Delta\sigma^z_i\sigma^z_j.
\end{aligned}
\label{Hamiltonianxxz}
\end{equation}
When $\Delta$=$1$ in Eq.~(\ref{Hamiltonianxxz}), Eq.~(\ref{Hamiltonian}) reverts to the ordinary Kondo-necklace model. We utilize iPEPS as the variational ansatz and optimize it for obtaining the ground state using automatic differentiation~\cite{PhysRevX.9.031041, Hasik}. Properties in the thermodynamic limit can be attained by exploiting the corner transfer matrix renormalization group (CTMRG)~\cite{doi:10.1143/JPSJ.65.891}. In this work, we choose the virtual bond dimension $D=4$ and the dimension for environment tensors $\chi=20$, which are found to be sufficient to obtain the qualitative phase diagram. For demonstration, we perform the bond dimension scaling analysis to check the stability of each phase (see Supplementary Note 1 and Supplementary Figure~\ref{figA2}).
Further information on the iPEPS method is provided in the Method section. Note that the interlayer coupling between $\tau$ field in Eq.~(\ref{Kondo-sieve}) is not included in Eq.~(\ref{Hamiltonian}).

In Fig.~\ref{fig1}(a), we demonstrate the zero-field phase diagram of 2D Kondo-necklace model ($\Delta$=$1$) with a pie chart. Starting from $\theta$=$0$, we obtain five different phases. In the antiferromagnetic (AFM) phase, the on-site $\boldsymbol{\sigma}$ and $\boldsymbol{\tau}$ moments are antiparallel, while nearest-neighbor $\boldsymbol{\sigma}$ ($\boldsymbol{\tau}$) moments also stay in antiparallel. With increasing $\theta$, ordering of both spin moments disappear and the system enters the spin-singlet phase through a continuous transition, with transition point at $\theta=0.31\uppi$ (recall that this is for $D=4$). 

After leaving the first quadrant, the first term in Eq.~(\ref{Hamiltonian}) favors ferromagnetic (FM) due to the sign change from plus to minus. With $\theta$ larger than 0.725$\uppi$ magnetic order continuously appears again but this time nearest-neighbor moments align in parallel, while $\boldsymbol{\sigma}$ and $\boldsymbol{\tau}$ remain antiparallel on each site. Thus we name this phase as $\sigma$-FM. 

Entering the third quadrant of the phase diagram, both nearest-neighbor and local coupling terms become FM, leading to the FM phase where all spins align in the same direction. Finally, in fourth quadrant the local coupling term is FM while nearest-neighbor coupling term becomes AFM. As a result, nearest-neighbor spin moments align antiparallel, but on-site moments are in parallel, becoming the $\sigma$-AFM phase. The phase transitions from $\sigma$-FM to FM, from FM to $\sigma$-AFM, and from $\sigma$-AFM to AFM are all of the first order since they break different translational symmetries. 

These phase transitions can be described by measuring the following order parameters: 
\begin{eqnarray}
\bar{S}  &=&\frac{1}{N}\sum_i\langle \boldsymbol{S}_i \rangle,\nonumber \\
\tilde{S}&=&\frac{1}{N}\sum_i (-1)^{i_x+i_y} \langle \boldsymbol{S}_i \rangle,\nonumber \\
S_-      &=&\frac{1}{N}\sum_i\langle \boldsymbol{\sigma}_i-\boldsymbol{\tau}_i \rangle
   = \langle\boldsymbol{\sigma}\rangle - \langle\boldsymbol{\tau}\rangle.
\label{Sorder}
\end{eqnarray}
Previous studies using an effective analytical approach (EAA)~\cite{PhysRevLett.80.5790}, quantum Monte Carlo (QMC)~\cite{PhysRevB.73.014431}, and stochastic series expansion (SSE)~\cite{PhysRevB.73.104450, IntJModPhys21.2245} obtained the transition point from AFM to spin-singlet at $K_\text{c}/J=1.37\approx \text{tan}(0.2993\uppi)$, $K_\text{c}/J=1.3888(1)\approx \text{tan}(0.3013\uppi)$, and $K_\text{c}/J=1.4\approx \text{tan}(0.3026\uppi)$ (at low $T$=$0.05J$), and they are indicated in Fig.~\ref{fig1}(b). Note that finite $D$ iPEPS tends to overemphasize the order parameter (see for example Figure 8 of the work by Hasik {\it et al.}~\cite{10.21468/SciPostPhys.10.1.012}) and thus, with higher $D$ we expect that the transition point predicted by iPEPS would get closer to the rest three.

\subsection{Field-induced BEC}

The spin-singlet ground state near $\theta$=$\frac{\uppi}{2}$ provides a good platform for the field-induced BEC after turning on the magnetic field. To see this, note that the elementary excitations over the singlet ground state are the mobile triplons carrying the quantum numbers $S^z$=$0, \pm 1$. With increasing field strength, the triplon band with $S^z$=$+1$ becomes more favorable and finally crosses the energy of the spin-singlet ground state, leading to the condensation of the $S^z$=$+1$ triplon. 
The local densities of triplons can be measured by the following operators:
\begin{align}
    \rho_i^\pm \equiv \frac{1}{2}\langle S_i^z(S_i^z\pm 1)\rangle,\quad
    \rho_i^0 \equiv \frac{1}{2} \langle \boldsymbol{S}_i^2 \rangle - \langle (S_i^z)^2 \rangle,
\label{rhoall}
\end{align}
where $\rho_i^{\pm}$ and $\rho_i^0$ are densities for the local triplon with $S_i^z$=$\pm 1$ and $S_i^z$=$0$ quantum numbers, respectively. The local density of the singlet is determined by 
\begin{eqnarray}
\rho_i^{\rm singlet}=1-(\rho_i^+ +\rho_i^0 +\rho_i^-)=1- \frac{1}{2}\langle\boldsymbol{S}_i^2\rangle.
\end{eqnarray}
Utilizing the triplon density operator $\rho^+_i$, we can define an order parameter as follows:
\begin{equation}
\begin{aligned}
    &\tilde{\rho}^+=\frac{1}{2N}\sum_i (-1)^{i_x+i_y} \langle S^z_i(S^z_i+1)\rangle,
\end{aligned}
\label{rhoplus}
\end{equation}
where $|\tilde{\rho}^+|>0$ reflects the translational symmetry breaking. For a complete characterization of the quantum phases, we define the condensate density order parameter that detects the $\text{U}(1)$ symmetry breaking:
\begin{equation}
    n_0\equiv \left|\frac{1}{N}\sum_i \langle B^\dagger_i \rangle \right|^2,
\label{n0}
\end{equation}
with $B^\dagger_i\equiv(-1)^{i_x+i_y}(\sigma^+_i-\tau^+_i)$. It is certain that $B_i^\dagger$ is non-trivial under the $\text{U}(1)$ symmetry. 

One can also see that its finite value reflects the mixture of the spin singlet and the triplon with $S_i^z$=$+1$, i.e., $|\langle B^\dagger_i \rangle| \propto \sqrt{\rho^+_i\rho^{\text{singlet}}_i}$ that occurs in the field-induced BEC. Recall that the magnetic field favors the $S^z$=$+1$ triplon band, causing the gap closing and eventually crossing with the singlet band.
This fulfills the general understanding of field-induced BEC that after band crossing, $S^z$=$+1$ band and singlet band are hybridized, resulting in BEC.

\begin{figure}[pbt]
\centering
	\includegraphics[width=1.\columnwidth]{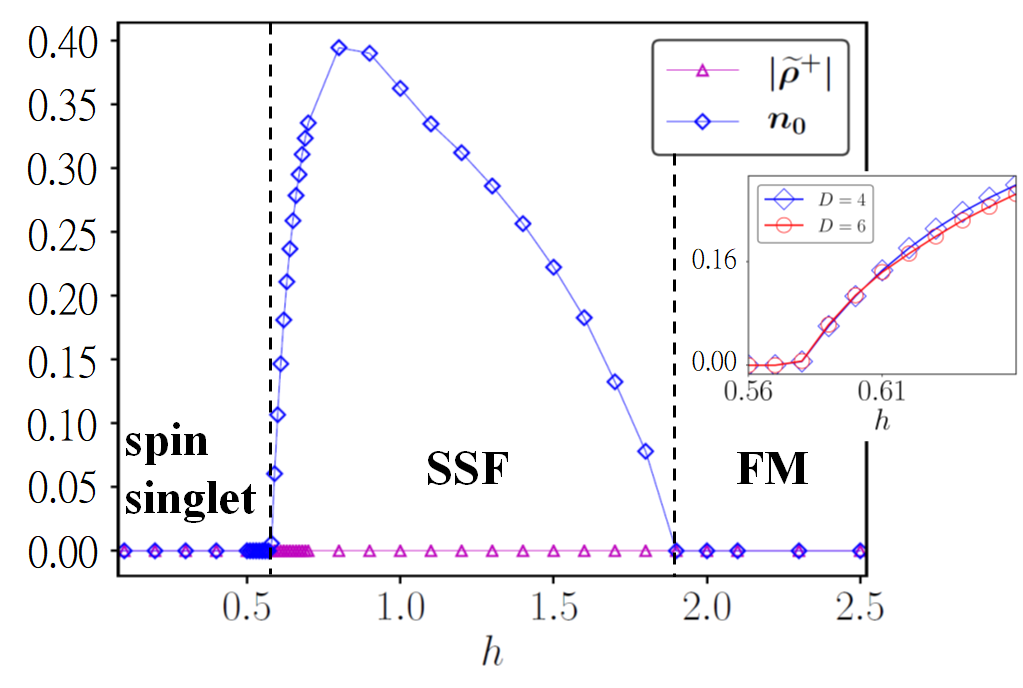}
\caption{Field-induced Bose-Einstein condensation for $H_{\text{KN}}$ with $\theta$=$0.4\uppi$. $h$ stands for the strength of field while SSF and FM represent the spin superfluid and ferromagnetic phases distinctly.
$|\tilde{\rho}^+|$ and $n_0$ come from the definition in Eqs.~(\ref{rhoplus}) and (\ref{n0}).
The small inset demonstrates $n_0$ for bond dimension $D=4$ and $D=6$ (dimension for environment tensors $\chi=36$) separately near the transition point from spin singlet to SSF. We can see that the transition point is located at around $h=0.58$ from both bond dimensions.
}
\label{fig2}
\end{figure}

Fig.~\ref{fig2} displays two order parameters, i.e., $\tilde{\rho}^+$ and $n_0$, as a function of the external field strength at $\theta$=$0.4\uppi$ ($K\approx 3J$). 
The spin-singlet phase remains stable before the ground state enters into the SSF phase around $h \approx 0.58$ at which the $S^z$=$+1$ triplon starts condensing. If we further enhance the magnetic field, finally spin moments are fully polarized and the asymptotic FM plateau appears, restoring $\text{U}(1)$ symmetry. 
Note that the triplon density remains uniform\,($\tilde{\rho}=0$), and both transitions are expected to be continuous due to the $\text{U}(1)$ symmetry breaking. 

\subsection{\label{SS} Combination of Field and Anisotropy}
\begin{table}
\centering
\begin{tabular}{cccccc}\hline\hline
            & \begin{tabular}[c]{@{}c@{}}~z-AFM~\\\end{tabular} & ~~SS~~ & ~~Solid~~ & ~~SSF~~ & ~~FM~~  \\ 
\hline
~Translation[$\widetilde{\rho}^{+}$]~ & X                                               & X  & X  & $\bigcirc$  & $\bigcirc$   \\ 
\hline
$\text{U}(1)$[$n_0$]       & $\bigcirc$                                               &  X  & $\bigcirc$  & X  & $\bigcirc$  \\
\hline\hline
\end{tabular}
\caption{The symmetry table for each phase.
z-antiferromagnetic, spin supersolid, spin superfluid, and ferromagnetic phases are abbreviated into z-AFM, SS, SSF, and FM separately.
$\bigcirc$ indicates the symmetry is present and X indicates it is broken. 
Orders that appear after translation and $\text{U}(1)$ symmetry breaking are also indicated in the brackets and their definitions come from Eqs.~(\ref{rhoplus}) and (\ref{n0}). 
Notice that the translational symmetry breaking denotes the appearance of spatial inhomogeneity in the triplon density, while the $\text{U}(1)$ symmetry breaking indicates the condensation of the $S^z$=$+1$ triplon.
z-AFM and Solid both possess $\text{U}(1)$ symmetry but have different quantum numbers $S^z$: zero for z-AFM while 0.5 for Solid. It is clear to see from the symmetry that a first-order transition takes place at the phase boundary between Solid and SSF, where phases break distinct symmetries on different sides.}
\label{tab1}
\end{table}

\begin{figure*}[tbp]
\centering
 	\includegraphics[width=2.0\columnwidth]{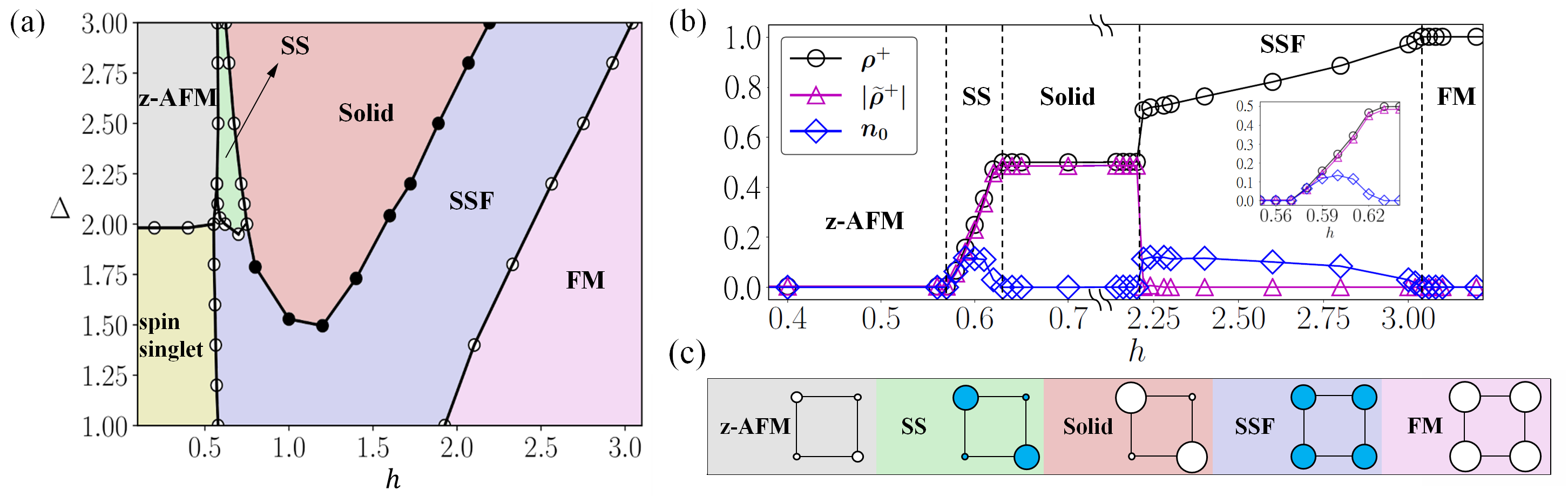}
\caption{The full phase diagram and detailed patterns. (a) The anisotropy-field phase diagram for 2D XXZ Kondo necklace model with $\theta$=$0.4\uppi$ ($J\cong0.31$ and $K\cong 0.95$), as defined in Eqs.~(\ref{Hamiltonian}) and (\ref{Hamiltonianxxz}). $h$ stands for the field strength and $\Delta$ represents the strength of XXZ anisotropy.
z-antiferromagnetic, spin supersolid, spin superfluid, and ferromagnetic phases are abbreviated into z-AFM, SS, SSF, and FM separately. The SS phase lies in the range $0.57 \lesssim h\lesssim 0.7$ with $\Delta\gtrsim 2$. Filled and empty circles indicate the first-order and second-order phase boundaries, separately. A more careful investigation for those continuous boundaries is non-trivial and thus we will leave it for the future works. (b) Order parameters along with $h$ for $\Delta$=$3$. Their definitions can be found in Eqs.~(\ref{rhoall}), (\ref{rhoplus}), and (\ref{n0}).
The inset shows the results obtained using bond dimension $D$=$5$ (dimension for environment tensors $\chi$=$25$), focusing on the SS area. 
(c) The triplon configurations for each phase. Within the $2\times 2$ unit cell, the size of each lattice site stands for the magnitude of $\rho^+_i$. Blue (white) color reflects nonzero (zero) condensation\,($n_0$). Here, with blue color it indicates $\text{U}(1)$ symmetry breaking.
Note that for z-AFM, $\rho^+_i$ has a checkerboard pattern with a small but nonzero on-site magnitude, and thus the homogeneity is broken and we have small but nonzero $\tilde{\rho}^+$ order.}
\label{fig3}
\end{figure*}

In the previous sections we have demonstrated the phase diagram for the 2D Kondo-necklace model in zero field and the field-induced BEC out of the spin-singlet phase due to the magnetic field.  
We expect a richer phase structure may emerge by introducing the XXZ anisotropy in the Heisenberg interaction. In the boson language, $\sigma_i^z \sigma_j^z$ is mapped to the repulsive interaction between neighboring bosons. 
While the interaction prefers the density wave or low density of triplons, the external field stabilizes dense populations of the $S^z$=$+1$ triplon. Indeed, we find that these competing effects give rise to various quantum phases, and the phase diagram is presented in Fig.~\ref{fig3}(a). 
To characterize the phases, we also show the order parameters in Fig.~\ref{fig3}(b) as a function of $h$ at $\Delta$=$3$.

In the absence of the field, the strong anisotropy results in a trivial magnetic state, named after z-antiferromagnetic (z-AFM) state, out of the spin-singlet phase where $\boldsymbol{\sigma}$ spins form the Néel configuration in $\hat{z}$ direction, and $\boldsymbol{\tau}$ spins align antiparallel to the on-site $\boldsymbol{\sigma}$ spin as well. The spin-singlet and z-AFM phases share the same $\text{U}(1)$ quantum number while the uniformity of the triplon density breaks seemingly continuously, suggesting the continuous transition\,(see Supplementary Note 1 and Supplementary Figure~\ref{figA1}(b)).
On the other hand, as increasing $\Delta$ or the repulsive interaction between triplons, the system evolves into a Solid phase out of the SSF phase in a wide region of the phase diagram. Note that the Solid phase is characterized by a checkerboard pattern of the $S^z$=$+1$ triplon density with a fractional number per unit cell, i.e., $\rho^+$=$1/2$ in Fig.~\ref{fig3}(b), which resembles the Mott phase for a fermionic system with large Hubbard $U$ interaction. Otherwise this is a factional magnetization plateau phase in the spin language. If we further enhance the strength of external field, the Solid state will melt down and SSF appears again. The first-order transition between Solid and SSF phases can be likely described by the XXZ model on the square lattice, which serves as the leading order mapping from the original Hamiltonian~\cite{PhysRevB.84.054432}.
It is worth noting that the $S^z$=$-1$ triplon is completely suppressed, i.e., $\rho^-$=$0$ throughout the phase\,(See Supplementary Note 1 and Supplementary Figure~\ref{figA1}(a)). We also find that a spin supersolid phase, which breaks both the uniformity of the triplon density and $\text{U}(1)$ symmetry simultaneously, appears in a narrow region indicated in Fig.~\ref{fig3}(a) and (b). The strong repulsive interaction introduced by XXZ anisotropy stabilizes the density wave of condensed triplons and thus gives rise to the translational symmetry breaking out of the SSF. The density wave is characterized by a $(\uppi,\uppi)$ wave vector. In order to visualize each phase, we illustrate the order parameters schematically in Fig.~\ref{fig3}(c), and summarize the symmetry properties of phases in Table~\ref{tab1}.

\section{\label{BNOAS}Relation to BNOAS}

Our work was stimulated by the report of the BNOAS insulator built on a $\text{d}^8$ Ni ion.
It was noted in the Introduction that the Kondo sieve model is the minimal spin Hamiltonian for BNOAS, which is three-dimensional (3D) but contains a layered structure. While one needs interlayer interaction for the 3D coupling, our 2D Kondo necklace model is expected to establish the underlying intralayer behavior. In addition, this is a good platform for study of the field-induced magnetic orders, besides the well-known examples $\text{TlCuCl}_3$ or $\text{SrCu}_2\text{(BO}_3\text{)}_2$. More interestingly, by introducing the symmetry required XXZ anisotropy the field-induced spin supersolid can be realized. Note that realization of supersolids has been an active research topic and recently reported to arise from a BEC made of dipolar atoms~\cite{PhysRevLett.122.130405, PhysRevX.9.011051, PhysRevX.9.021012, PhysRevLett.123.050402, Nature.574.382, Nature.574.386}. However, such realization with the cold-atom equipment is not an easy task and therefore we do not have many examples so far, considering the date when this concept of supersolid was first proposed~\cite{PhysRev.104.576}. 

As suggested by Ng and Lee~\cite{PhysRevLett.97.127204}, instead of searching for a supersolid phase at very low temperature, magnetic materials with spin singlets in their ground states provide a more promising scenario for its formation. For BNOAS, the ``local'' spin and ``Kondo'' spin moments both arise from electrons in $\text{e}_\text{g}$ subshell of Ni. Earlier work had revealed that the Ni ion in 2D materials is suitable for the study of XXZ-type antiferromagnetism~\cite{PhysRevB.46.5425, PhysRevB.92.224408, ncomm.10.345}. These discoveries indicate that BNOAS has the potential to serve as a good platform for the realization of a spin supersolid phase, making it worthwhile to study BNOAS further, theoretically and experimentally, and exploit more of its underlying physics. 

As mentioned by Jin {\it et al.}~\cite{PhysRevResearch.2.033197}, doping electrons into BNOAS leads towards the region of possible high-$\text{T}_\text{c}$ superconductivity, considering the similarities to hole-doped cuprates and Sr-doped $\text{NdNiO}_2$ superconductors~\cite{nature.572.624}. In addition, the spin-singlet state serves as another scenario -- an unusual one -- for a self-doped Mott insulator~\cite{PhysRevB.101.020501}. For $\text{Nd}_{1-x}\text{Sr}_x\text{NiO}_2$, the sparse Nd $5\text{d}$ conduction carriers may couple with Ni $3\text{d}_{\text{x}^2-\text{y}^2}$ electrons to form Kondo singlets dynamically. Such singlets will suppress the AFM order, leading to a paramagnetic ground state which can be metallic. For BNOAS, on the other hand, without doping we have a magnetically inert ground state with a Kondo singlet occupying every site. Because of the hard-core nature of the on-site singlets, it should be an insulator. Upon doping BNOAS, however, the singlet density decreases while the long-range AFM order is hindered unless a high doping level. Therefore, with an intermediate doping level we suggest that a insulator-metal transition could also take place. 

Extending these similarities, this self-doped superconducting transition can be modeled by a $t-J$-like Hamiltonian~\cite{PhysRevB.101.020501}. Here we propose an effective Hamiltonian for describing the microscopic mechanism for electron-doped BNOAS. We start from the $3\text{d}_{\text{x}^2-\text{y}^2}$ spins ($\boldsymbol{\sigma}$) and adopt an anisotropic $t-J$ model
\begin{equation}
\begin{aligned}
H_\text{J}=&-t\sum_{\langle ij \rangle \alpha}\text{P}_\text{G}(c^\dagger_{i\alpha}c_{j\alpha}+\text{H.C.})\text{P}_\text{G}+J\sum_{\langle ij \rangle}(\boldsymbol{\sigma}_i\cdot\boldsymbol{\sigma}_j)_{\Delta},
\end{aligned}
\label{HamiltonianJ}
\end{equation}
where $c_{i\alpha}$ ($c^\dagger_{i\alpha}$) is the annihilation (creation) operator for $3\text{d}_{\text{x}^2-\text{y}^2}$ electrons with $\alpha=\uparrow,\downarrow$ and H.C. denotes the Hermitian conjugate. The spin operator is connected to fermionic operator by $\sigma^\beta_i=\frac{1}{2}\sum_{\alpha,\alpha'}c^\dagger_{i\alpha}\rho^\beta_{\alpha,\alpha'}c_{i\alpha'}$ where $\rho^\beta$ is the Pauli matrix with $\beta$=$x,y,z$. $\text{P}_\text{G}$=$\Pi_i(1-n_{i\uparrow}n_{i\downarrow})$, with $n_{i\alpha}$=$c^\dagger_{i\alpha}c_{i\alpha}$, is the Gutzwiller projection operator to prevent the double occupancy on each site~\cite{PhysRev.134.A923}. Longer-range hopping can be also included to better explain experimental observations but here we only demonstrate the nearest-neighbor hopping. While the $t-J$ model is often applied for hole-doped cuprate superconductors, its electron-doped counterpart only requires a sign change for the hopping constant due to the particle-hole transformation and thus we can directly borrow the same form here~\cite{PhysRevB.49.3596}. Additionally, to preserve the degree of freedom for the XXZ anisotropy, we retain the form of $($ $\cdot$ $)_{\Delta}$ for the superexchange term.

The $3\text{d}_{\text{z}^2}$ spin $\boldsymbol{\tau}$ is coupled to the $\text{d}_{\text{x}^2-\text{y}^2}$ spin $\boldsymbol{\sigma}$ and can be described by $H_\text{K}$:
\begin{equation}
\begin{aligned}
H_\text{K}=K\sum_{i}\vec{\sigma}_i\cdot\vec{\tau}_i.
\end{aligned}
\label{HamiltonianK}
\end{equation}
The final effective Hamiltonian is the sum 
$H_{\text{eff}}=H_\text{J}+H_\text{K}$. This Hamiltonian, which we call the $t-J-K$ model, can be numerically solved with various methodologies~\cite{PhysRevLett.92.227002, PhysRevB.78.134530, PhysRevB.84.041108, PhysRevLett.113.046402, scirep.6.18675, NewJPhys.19.013028, scirep.9.1719} by Monte Carlo~\cite{RevModPhys.73.33} or renormalized mean-field theory~\cite{springerthesis} besides tensor network, for its properties in real and momentum space. Thus, we consider this to be our future challenge.

\section{\label{summary}Summary}

In this work, we make use of iPEPS, a 2D tensor network ansatz, to solve the Kondo necklace model in two dimensions. Without the XXZ anisotropy, we obtain the zero-field phase diagram and locate the region of spin-singlet formation. Upon turning on an external magnetic field, the spin-singlet phase goes through a phase transition into spin superfluid, a well-known phenomenon called field-induced BEC. By adding XXZ anisotropy, we argue that now the triplet state with $S^z$=$0$ is more favorable and closes the gap with large enough anisotropy $\Delta$. With external field, an exotic spin supersolid phase appears between two magnetization plateaus. We provide a $\Delta$-$h$ phase diagram and relocate the region where we believe spin supersolid can be realized with 2D Kondo-necklace model. 

Since BNOAS has (weakly) coupled infinite-layer nickelate planes, and adopting the 3D Kondo sieve model to be its effective Hamiltonian, we expect such field-induced BEC/spin supersolid phases can be realized within its nickel-oxide layer. For a more careful investigation in the future we will consider probing the continuous phase boundaries in more detail for their universality class or critical exponents. Moreover, QMC provides another useful tool, especially for finite temperature. Such thermalization can also be approached by iPEPS through a purification process and it shows good consistency with QMC~\cite{PhysRevResearch.1.033038}. It will be of great interest to study the thermal properties of Kondo-necklace model with both numerical techniques and thus one of our future works.

Looking toward extending theoretical work, we propose an effective $t-J-K$ Hamiltonian that can be used to describe the potential superconductivity arising in BNOAS after doping. Further studies for this effect from both experimental and theoretical sides are again of great interest and regarded as our future goal, too.

\section{\label{Acknowledgement}Acknowledgement}
W.-L.T. thanks the computational resources provided by the Kawashima research group from the Institute for Solid State Physics (ISSP), University of Tokyo. Many important inspirations were triggered during the \textsl{International Workshop on Quantum Magnets in Extreme Conditions} organized by ISSP on March 22-26, 2021.
This work was supported by National Research Foundation (NRF) of Korea under the grant numbers NRF-2020R1I1A3074769\,(H.-Y.L. and W.-L.T.), NRF-2019R1A2C1009588\,(K.-W.L.) and NRF-2020R1A4A3079707, NRF-2021R1A2C4001847\,(E.-G.M.). H.-Y.L. was also supported by Basic Science Research Program of NRF funded by the Ministry of Education (MOE) of Korea\,(2014R1A6A1030732). This research was partially supported by the Fostering Outstanding Universities for Research (BK21 FOUR) project funded by the MOE and NRF of Korea.
W.E.P. acknowledges supported from National Science Foundation Grant No. DMR 1607139.

\section{\label{method}Methods}
\begin{figure}[!hbt]
\centering
	\includegraphics[width=1.0\columnwidth]{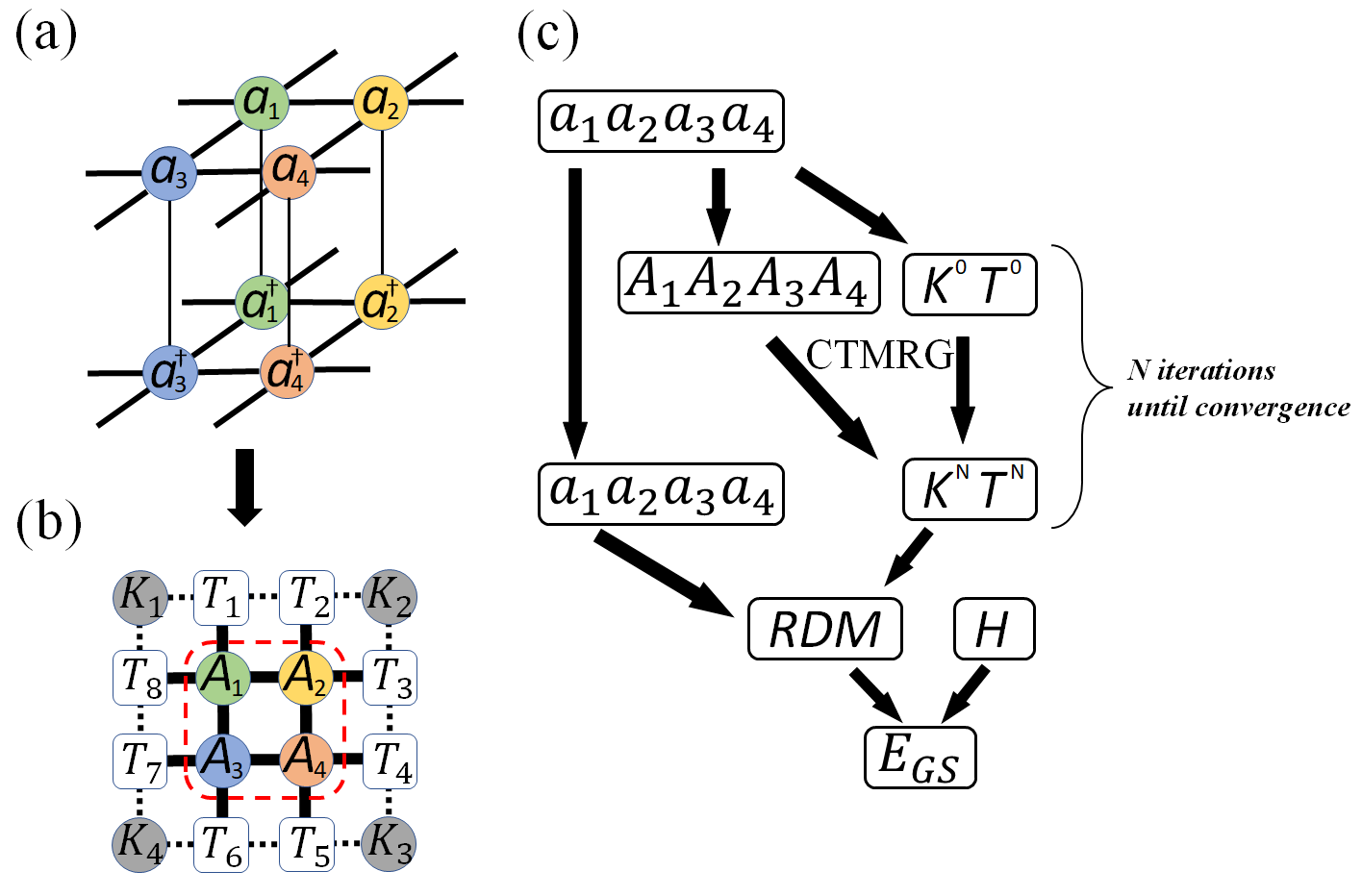}
\caption{Some basics for the infinite projected entangled-pair state ansatz. (a) The four bulk tensors $a_1$, $a_2$, $a_3$, and $a_4$, with bond dimension $D$ for each bond. After tracing out the physical bonds with their complex-conjugate tensors ($a_1^\dagger$ to $a_4^\dagger$), we obtain the double-layered tensors $A_1$, $A_2$, $A_3$, and $A_4$ shown in (b). Along with eight edge tensors ($\textsl{T}_1$ to $\textsl{T}_8$) and four corner tensors ($\textsl{K}_1$ to $\textsl{K}_4$), the norm of ansatz, $\langle \Psi|\Psi\rangle$, can be calculated as the tensor graph in (b). The black thick bond is of dimension $D^2$ while dotted bonds have dimension $\chi$. Area enclosed by red dashed line is the unit cell. In (c) we demonstrate the computation graph for the GS energy $E_{\text{GS}}$ from the initial input to be the bulk tensors. CTMRG stands for the corner transfer matrix renormalization group and RDM refers to the reduced density matrix. H is our target Hamiltonian.}
\label{fig4}
\end{figure}
\subsection{Infinite Projected Entangled-pair State}
In this work we adopt the 2D tensor network ansatz, the infinite projected entangled-pair states, for our calculation. The iPEPS tensor network ansatz is an effective numerical method for dealing with various quantum problems in two dimensions~\cite{PhysRevLett.101.250602, PhysRevB.84.041108, PhysRevLett.113.046402, NatRevPhys.1.538}. It has many merits that some other numerical techniques do not have, such as its attainability to the thermodynamic limit and freedom from the vicious sign problem in quantum Monte Carlo. This ansatz contains two parts, the bulk tensors and environmental tensors for achieving the infinite size. Technically, the number of bulk tensors can be freely chosen, but a better choice will allow it to be able to reflect the real space modulation for the ground state (GS). Thus, as shown in Fig.~\ref{fig4}(a), in this work we apply a $2\times 2$ unit cell for the bulk (we have also tried other bulk sizes, see Supplementary Note 2), meaning that there are four different rank-5 bulk tensors from $a_1$ to $a_4$. For each tensor it has four virtual bonds (bonds that connect nearby tensors), capturing the entanglement between site to site, with bond dimension $D$. The remaining one leg of each tensor, connecting $a_n$ and $a_n^\dagger$, is the physical bond with dimension $d$, equal to the dimension of local Hilbert space. The accuracy of iPEPS method is determined by the bond dimension since larger $D$ captures better the entanglement among sites. For critical systems we usually need a larger $D$ since the correlation length diverges. However, for quantum states that are less entangled, such as the scenario in this work, we have finite correlation length and thus the results remain nearly the same after an adequate $D$.

With bulk tensors, the iPEPS ansatz is complete with the environmental tensors in Fig.~\ref{fig4}(b), where rank-2 $\textsl{K}_n$ tensors stand for the corners and rank-3 $\textsl{T}_n$ tensors for the edges ($n=1$ to 4). $A_1$ to $A_4$ tensors correspond to double-layered tensors by tracing out the physical bonds, as indicated in Fig.~\ref{fig4}(a). The dimension for \textsl{K} tensors is $\chi\cdot\chi$ while for \textsl{T} tensors it is $D^2\cdot\chi\cdot\chi$. The environmental tensors can be obtained from double-layered tensors through a process called CTMRG~\cite{doi:10.1143/JPSJ.65.891, PhysRevB.80.094403, PhysRevLett.113.046402}. With corner and edge tensors, not only can we extrapolate the system size to the infinity, but the physical observables can also be calculated by constructing the corresponding reduced density matrix.

\subsection{Tensor optimization}
With the structure of iPEPS explained above, next we discuss how to optimize this ansatz. Since the environmental tensors come from bulk tensors, the number of variational parameters is decided by the bulk. Thus, optimization of iPEPS concerns the optimization of the bulk tensor. Traditionally, people make use of the technique called simple or full update based on the imaginary-time evolution for the optimization~\cite{Ran2020}. On the other hand, as a variational ansatz, a direct minimization of GS energy by changing variational parameters systematically based on the gradients might be a more direct approach. Nonetheless, it is not an easy job to evaluate the energy gradient of each variational parameter~\cite{PhysRevB.94.035133}. 

Recently, a new way of calculating such energy gradients has been proposed by using the technique called automatic differentiation~\cite{PhysRevX.9.031041}. It is a numerical way of evaluating function derivatives to machine precision and is often applied for updating the neural network for machine learning~\cite{Bartholomew2000, Baydin2018}. The idea of automatic differentiation is based on the chain rule: it assumes that a numerical function is composed of elementary operations (addition, subtraction, multiplication, and division) and functions ($\mathrm{sin}$, $\mathrm{log}$, etc.), irrespective of the complexity. For a given function, we can visualize its computation process by constructing a computation graph composed of individual nodes, where each node represents a intermediate result. The inputs will go through the computation graph and produce the final output. Later, by applying the forward or backward propagation, we are able to obtain the derivatives of the outputs with respect to each input.

\subsection{Working Flow}
Since our problem here has multiple inputs (tensor elements as the variational parameters) and only one output (GS energy, $E_{\text{GS}}$), it is more natural to apply the backward mode. First we need to record down the computation graph from initial bulk tensors to the final energy, as demonstrated in Fig.~\ref{fig4}(c). Starting from bulk tensors, we can construct double-layered tensors and initial environmental tensors ($\textsl{K}^0$ and $\textsl{T}^0$). By applying CTMRG until convergence, we obtain the effective final environment with $\textsl{K}^N$ and $\textsl{T}^N$. Then we construct the reduced density matrix with bulk tensors and environment. Finally, along with the Hamiltonian, we can compute $E_{\text{GS}}$. After completing one computation flow (one epoch), the energy gradients are evaluated by backward propagation, and we make use of these gradients to update the tensor elements with a desired degree (learning rate). After a large enough number of epochs, the energy converges and we obtain a good GS ansatz for further calculation of physical observables, or for other investigations of the mother Hamiltonian. Please refer to the open repository by Hasik {\it et al.}~\cite{Hasik} for a practical package of this iPEPS method with automatic differentiation.

\section{\label{data}Data availability}
The authors declare that the main data supporting the findings of this study are available within the article and its Supplementary Discussion. All relevant data in this paper are available from the authors upon reasonable request. An open access repository for the basic codes of iPEPS and datasets is available at https://doi.org/10.5281/zenodo.6420017.




\bibliography{draft}
\clearpage

\title{Field-induced Bose-Einstein condensation and supersolid in the two-dimensional Kondo necklace: Supplementary information}

\maketitle

\section{\label{triplon}Supplementary Note 1: Triplon densities}
Here, we present the triplon densities as a function of the field strength $h$ at $\Delta$=$3.0$. The triplon densities can be measured as follows:
\begin{align}
&\rho^{\pm} \equiv \frac{1}{2N}\sum_{i=1}^N \langle S_i^z(S_i^z \pm 1) \rangle,\\
&\rho^{0} \equiv \frac{1}{2N}\sum_{i=1}^N \langle \boldsymbol{S}_i^2 \rangle - \langle (S_i^z)^2 \rangle,\\ 
&\rho^{\rm singlet} \equiv 1 - (\rho^+ + \rho^0 + \rho^- ).
\label{eq1}
\end{align}
Note that the densities do not change in the z-AFM and Solid phases while increasing the strength of the magnetic field. That is because the $\text{U}(1)$ symmetry is preserved, and thus the averaged density of each triplon remains constant. It is worth noting that the density of $S^z$=$-1$ tripon, $\rho^-$, is completely suppressed in the Solid phase while the one of $S^z$=$+1$ triplon is $1/2$, indicating that the Solid phase is a magnetization plateaux phase with a fractional magnetization $m^z$=$\rho^+ - \rho^-$=$1/2$. The results are demonstrated in Supplementary Fig.~\ref{figA1}(a). For the transition of spin singlet to z-AFM, we demonstrate the $|\tilde{\rho}^+|$ to $\Delta$ plot in Supplementary Fig.~\ref{figA1}(b). A continuous phase transition is indicated with our numerical results. However, due to the difficulty of examining the phase boundary with finite $D$, we leave the further investigation for the nature of this phase transition for the future works.

In Supplementary Fig.~\ref{figA2}(a), we extrapolate the two order parameters, $|\tilde{\rho}^+|$ and $n_0$ at $h=0.6$, to infinite $D$. We obtain the results that $|\tilde{\rho}^+|_{D\rightarrow\infty}=0.2202$ and ${n_0}_{D\rightarrow\infty}=0.1236$, suggesting that these two orders will remain coexisting for spin supersolid phase. We also make the similar extrapolation of $\rho^+$ for Solid and SSF phases in Supplementary Fig.~\ref{figA2}(b). The result shows consistency after increasing $D$ and for SSF (Solid) phase $\rho^+_{D\rightarrow\infty}$=$0.1212$ (0.5).
\begin{figure}[!hbt]
\centering
	\includegraphics[width=1.0\columnwidth]{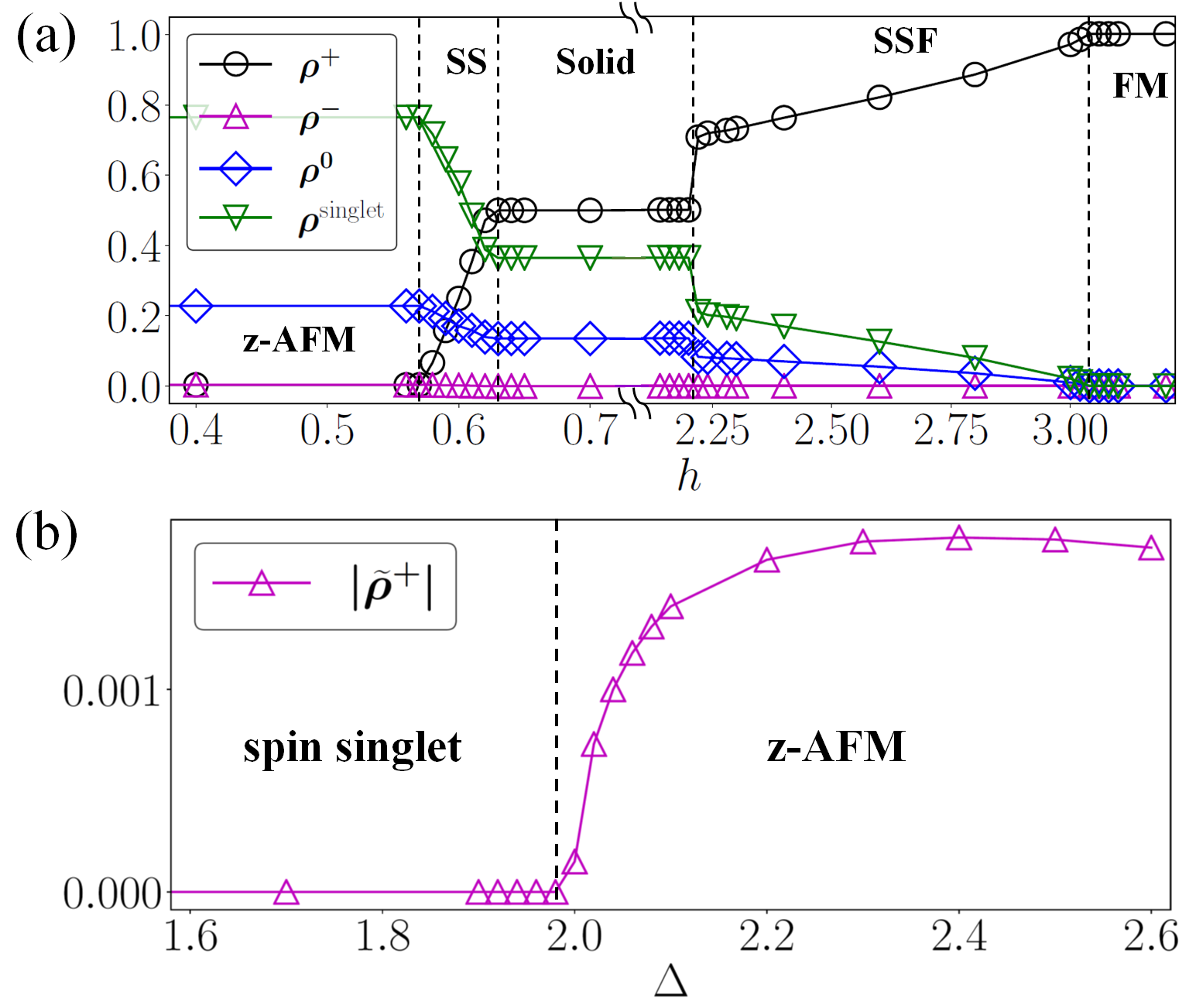}
\caption{Detailed orders by iPEPS calculation. (a) Plot of the triplon densities as a function of the strength of the magnetic field ($h$) at the XXZ anisotropy $\Delta$=$3.0$. z-antiferromagnetic, spin supersolid, spin superfluid, and ferromagnetic phases are abbreviated into z-AFM, SS, SSF, and FM separately. Definitions of corresponding orders can be found in Eqs. (5) and (6) in the Main Article. (b) Plot of $|\tilde{\rho}^+|$ (defined in Eq. (7) of the Main Article) to $\Delta$ (the strength of XXZ anisotropy) at $h=0$.}
\label{figA1}
\end{figure}
\begin{figure}[!hbt]
\centering
	\includegraphics[width=1.0\columnwidth]{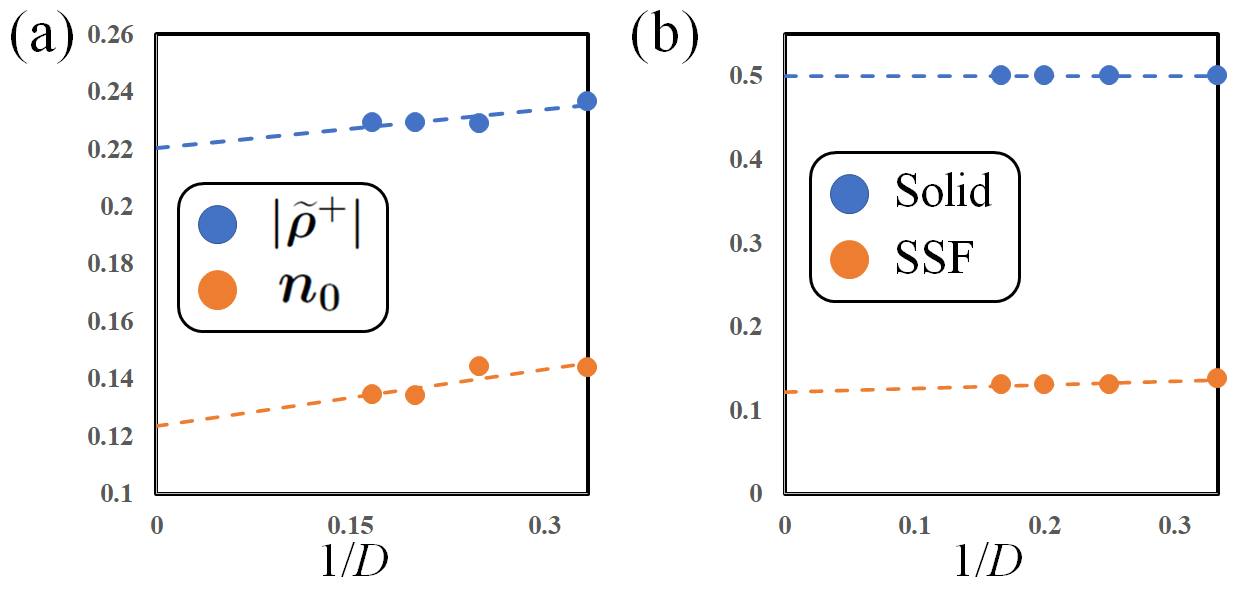}
\caption{(a) The extrapolation to infinite bond dimension ($D$) of order parameter $|\tilde{\rho}^+|$ (defined in Eq. (7) of the Main Article) and condensate density $n_0$ (defined in Eq. (8) of the Main Article), for $h=0.6$ (the field strength). (b) The same extrapolation of $\rho^+$ (defined in Eq.~(\ref{eq1}) in the Supplementary Note 1) for Solid ($h$=$1.5$, $\Delta$=$3.0$) and spin superfluid (SSF) ($h$=$0.65$, $\Delta$=$1.0$) phases. $\Delta$ stands for the the strength of XXZ anisotropy.}
\label{figA2}
\end{figure}
\begin{table}
\centering
\begin{tabular}{cccccc}\hline\hline
            & \begin{tabular}[c]{@{}c@{}}~~2$\times$2~~\\\end{tabular} & ~~3$\times$2~~ & ~~4$\times$2~~ & ~~3$\times$3~~  \\ 
\hline
$h=0.4$(z-AFM) & -0.837(6)                                               & -0.79(1)  & -0.837(6)  & -0.82(0)     \\ 
\hline
$h=0.6$(SS)       & -0.841(1)                                               &  -0.80(2)  & -0.841(1)  & -0.79(2)    \\
\hline
$h=1.5$(Solid)       & -1.288(3)                                               &  -1.18(2)  & -1.288(2)  & -1.17(6)    \\
\hline
$h=2.5$(SSF)       & -1.863(8)                                               &  -1.858(0)  & -1.863(6)  & -1.843(1)    \\
\hline\hline
\end{tabular}
\caption{The energy table of different unit cells for iPEPS calculation. z-antiferromagnetic, spin supersolid, and spin superfluid are abbreviated into z-AFM, SS, and SSF separately and $h$ denotes the field strength.
The last digit of each number in the parenthesis indicates the error in the last digit. Error digits are determined by checking the rate of convergence in the last few steps of the iteration of our variational optimization.}
\label{tabA1}
\end{table}

\section{\label{unit cells}Supplementary Note 2: Effect of Different Unit Cells}
Since the calculations in the Main Article are all executed using a 2$\times$2 unit cell, it is of interest to probe the calculation with different unit cells and see whether there can be other stable solutions or even a new magnetization plateau. For this purpose, we pick one point from each phase in Supplementary Fig.~\ref{figA1}(a) and compare the energies obtained by iPEPS under different unit cells. The results are demonstrated in Supplementary Table~\ref{tabA1}. We can easily find out that for the 2$\times$2 and 4$\times$2 unit cells they share very close energies and are always favored. It is because they are both commensurate to the minimum unit cell required for our antiferromagnetic Hamiltonian. On the other hand, for 3$\times$2 or 3$\times$3 unit cell not only the energies are higher but also the accuracy is poor for those translation-symmetry-breaking phases. This effect is because of the incommensurability, which causes a very slow or even unconvergent CTMRG process. As a result, we argue that for our 2D Kondo-necklace model in the square lattice, due to the lack of frustration, a 2$\times$2 unit cell is the most suitable and efficient for studying its behaviour.




\end{document}